\documentclass[]{aa}
\usepackage{natbib}
\usepackage{graphicx,amsbsy}
\usepackage{epsfig}
\newcommand{\teff}{$T_{\rm eff}$}
\newcommand{\logg}{$\log g$}
\newcommand{\feh}{[Fe/H]}

\newcommand{\ebv}{$E(B-V)$} 
\newcommand{\eby}{$E(b-y)$} 
\newcommand{\evy}{$E(v-y)$}

\def\farcs{\hbox{$.\!\!^{\prime\prime}$}}
\def\farcm{\hbox{$.\mkern-4mu^\prime$}}

\def\arcsec{\hbox{$^{\prime\prime}$}}

\def\logg{\hbox{$\log g$}}

\def\logg{log\hspace*{1mm}$g$}

\begin{document}
\authorrunning{Willemsen et al.}
\titlerunning{Analysis of medium resolution spectra}
\title{Analysis of medium resolution spectra by automated methods  - application to M\,55 and $\omega$ Centauri\thanks{Based
on observations obtained at the European Southern Observatory, Chile
(Obs.
Prog. 67.D-0300 and 69.D--0172)}}

\author{ P.G. Willemsen\inst{1},
    \ M. Hilker\inst{1},
    \ A. Kayser\inst{1,2},
    \ C.A.L. Bailer-Jones\inst{3}
       }

\offprints{\tt{willemse@astro.uni-bonn.de} }

\institute{ Sternwarte der Universit\"at Bonn, Auf dem H\"ugel 71, 53121
  Bonn, Germany \and
  Astronomisches Institut der Universit\"at Basel, Venusstrasse 7, 4102 Binningen, Switzerland
  \and Max-Planck-Institut f\"ur Astronomie, K\"onigstuhl 17, 69117 Heidelberg, Germany
}

\date{}

\abstract{We have employed feedforward neural networks trained
on synthetic spectra in the range 3800 to 5600 \AA\ with
resolutions of $\sim$ 2-3 \AA\ to determine metallicities from spectra
of about 1000 main-sequence turn-off, subgiant and red giant stars in
the globular clusters M\,55 and $\omega$ Cen. The overall metallicity
accuracies are of the order of 0.15 to 0.2 dex.  In addition, we
tested how well the stellar parameters \logg\ and \teff\ can be
retrieved from such data without additional colour or photometric
information. We find overall uncertainties of 0.3 to 0.4 dex for
\logg\ and 140 to 190 K for \teff .  In order to obtain some measure
of uncertainty for the determined values of \feh , \logg\ and \teff ,
we applied the bootstrap method for the first time to neural networks
for this kind of parametrization problem.  The distribution of
metallicities for stars in $\omega$ Cen clearly shows a large spread
in agreement with the well known multiple stellar populations in this
cluster.

}

\maketitle

\section{Introduction \newline} 

The advance in modern spectroscopic techniques coupled with 8-meter
class telescopes makes it possible to simultaneously obtain spectra
for a large number of stars in clusters. Although such spectra as
analysed in this work do not have high enough resolution to make
detailed line analyses, they offer the possibility to measure
abundances with good precision ($\sim$ 0.2 dex). This permits, for
example, statistically significant analyses of stellar populations in
dwarf spheroidal galaxies or studies of the intra- and intercluster
abundance variations of globular clusters.

While the data acquisition and subsequent reduction are themselves
challenging, the analysis of such a large amount of data can be
efficiently handled by the introduction of automated methods. This can
be as accurate as spectral analyses (using line indices etc.) carried
out by a human expert, as proven in several earlier works (see
e.g. \citealt{Snider2001}, \citealt{Soubiron98}).  Several of these
works show that the accuracy of the results was limited
by the quality of the data rather than by the automated algorithm used to
analyse them.
 
 \begin{figure*}[t]
\centering
\includegraphics[scale=0.9]{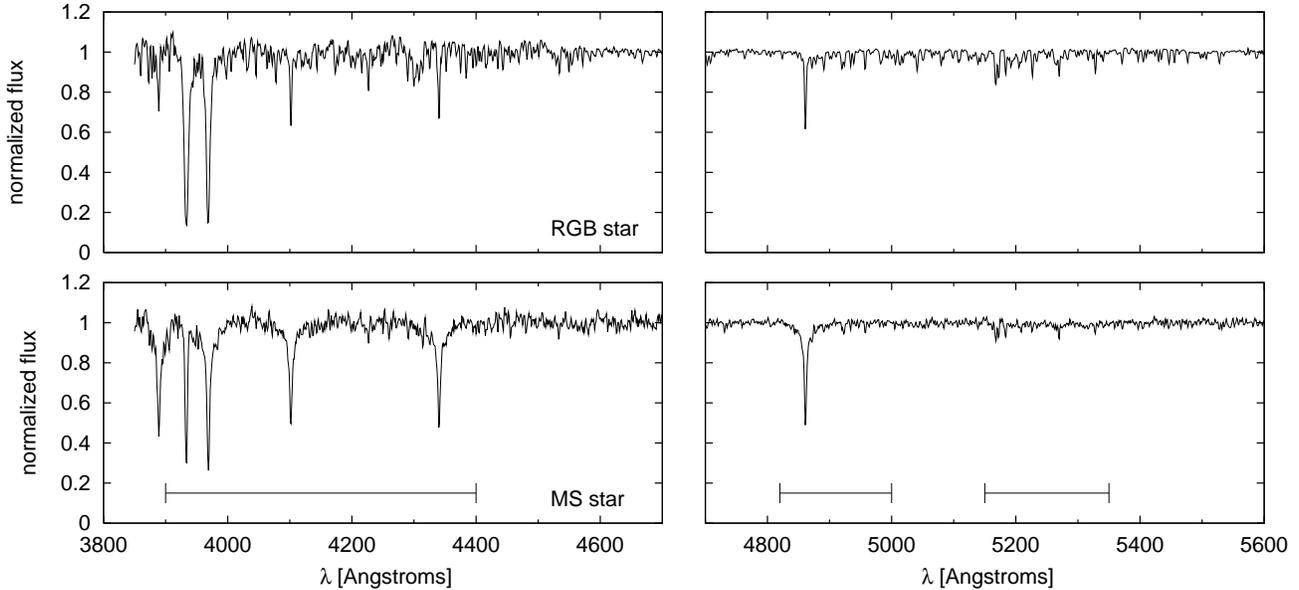}
\caption{Examples of observed spectra. The left panels show the
spectra as obtained by the blue grism, the right column that of the
red grism. The upper panel shows spectra of a red giant star, the
lower for a main sequence star in M\,55, i.e. \feh\ = $-$1.8 dex. Note
the larger number of absorption lines in the spectrum of the cool
giant. In the lower plot the wavelength ranges used for the neural network analysis are indicated.
\label{spectra}}
\end{figure*}
 
Recent studies using automated methods to retrieve stellar parameters
\teff , \logg\ and \feh\ from observed data were limited in various
ways. (In the following we call this process `parametrization'.) For
example, \cite{Bonifacio03} employed a minimum distance-based
classifier to high resolution spectra of stars in the
Sagittarius Dwarf Spheroidal but were restricted to certain
luminosities and temperatures.  Earlier efforts, especially
those using neural networks, were also limited in their analysis of
errors of the determined stellar parameters. Most often, the
precision with which stellar parameters can be retrieved is evaluated
by using a validation set with known target values. While this is correct to get some estimate of the overall uncertainty,
this technique cannot be applied to previously unseen data, where such
additional information is of course not available. However, any
parameter estimate is only meaningful if some direct measure of
uncertainty can be applied.  To overcome this, we employed the
bootstrap method to neural networks which allows us to assign
individual error bars to the estimated stellar parameters.

\cite{Snider2001}, using neural networks on spectral data with
resolutions similar to those in this work, could determine all three
stellar parameters (\teff , \logg\ and \feh) well. Indeed, they showed
that the precisions are similar to those that can be obtained from
fine analyses of high resolution spectra. While their neural network
training and validation sets contained spectra from spectroscopic
observations, our work relies entirely on synthetic spectra as
templates. Of course, `real world' spectroscopic data with stellar
parameters assigned from spectral synthesis analyses are the ideal
case but are expensive in terms of
observing time and effort. In this work we demonstrate that synthetic
spectra in combination with automated methods can be used to determine
reliable stellar parameters from spectroscopic observations.

We examine spectra from stars in the globular clusters M\,55 and
$\omega$ Centauri. These clusters serve as a demonstration of
our methods. We have recently obtained a much larger database of
stellar spectra in several globular clusters, and these will be
subject to an extensive analysis in the future. M\,55 has a
metallicity of about $-$1.80 dex (\citealt{Zinn84},
\citealt{Harris96}, \citealt{Richter99}) and is used primarily to
validate our method.  $\omega$ Cen, in contrast, is known to harbour
several stellar populations with different kinematical characteristics
and a metallicity range from $-$2.5 to $-$0.5 dex (see e.g
\citealt{Norris95}, \citealt{Suntzeff96}), suggesting an entirely
different formation history to that of normal globular
clusters. Indeed, several earlier spectroscopic and photometric
studies found a prolonged formation period of about 2 to 5 Gyr,
suggesting that this object is more likely the nucleus of an accreted
dwarf galaxy than a globular cluster (\citealt{Lee99},
\citealt{Hilker20}, \citealt{Hughes04}, and especially
\citealt{Hilker04}). In addition, it was found that there is a
clear trend of the $\alpha$ element abundances with the overall
metallicity (see e.g. \citealt{Pancino02} or \citealt{Origlia03}).
The populations in $\omega$ Cen can be subdivided into a dominant
metal-poor component with a metallicity peak at $-$1.7 dex which
accounts for about 70\% of the stars. Some 25\% of the stars belong
to the intermediate metallicity population with \feh\ $\sim$ $-$1.2
dex while the last 5\% are metal-rich stars with \feh\ $\sim$ $-$0.6
dex (see e.g. \citealt{Smith04}).

The purpose of this work is to demonstrate the ability of
automated methods (here neural networks) to derive accurate
metallicities as well as surface temperatures and gravities from
spectra with medium resolutions of $\sim$ 2-3 \AA . Sect.\,\ref{dataq}
and \ref{obsspec} give a summary of the acquisition, reduction and
selection of the observations, while Sect.\,\ref{syndata} describes
the set of synthetic spectra. The training procedure of the neural
networks and the application of the bootstrap method are described in
Sect.\,\ref{networks}. In Sect.\,\ref{modelval} the networks are
validated on (observed) spectra of stars in M\,55 (with known
metallicity and \logg\ and \teff\ estimated from isochrones) as well
as specific stars with known stellar parameters. The
same networks are then applied to stellar spectra in $\omega$ Cen in
order to derive preliminary metallicity values. Since this cluster is
indeed very peculiar, we do not attempt to interpret these results in
great detail. Rather, we want to show that the networks can yield
reliable results across a large range of abundances for stars in
different evolutionary states.

\begin{figure}[h]
\centering
\includegraphics[scale=0.38,angle=-90]{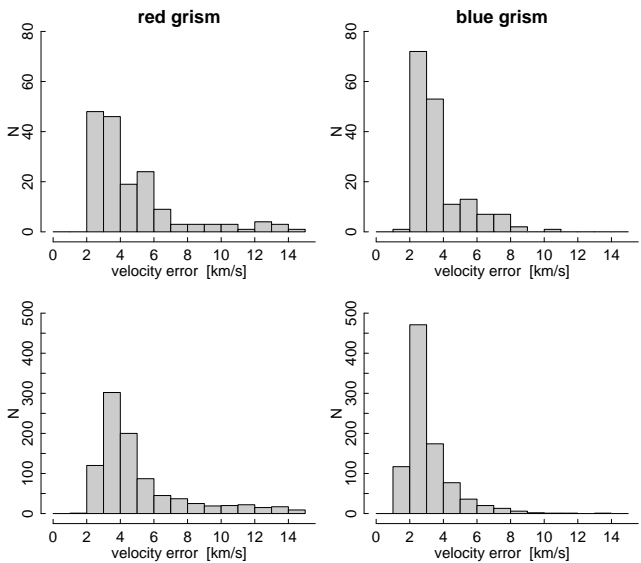}
\caption{Distributions of the velocity errors of the red grism (left
column) and blue grism (right column) for the spectra of M\,55 (top
panels) and $\omega$ Cen (bottom panels).  The errors in the red grism
are somewhat larger because of a smaller number of prominent absorption
lines available for the cross-correlation procedure.
\label{histerr}}
\end{figure}

\section{Observations and reductions}
\label{dataq}

\subsection{Data acquisition}

The data were obtained on the nights 7-12 May 2002 at the VLT at
ESO/Paranal (Chile) in visitor mode. The instrument was the FORS2 low
dispersion spectrograph in combination with the mask exchange unit
(MXU) and a 4 $\times$ 4 k MIT CCD with a field of view of 6.8 square
arcminutes.  On average, about 60 stars were observed per
mask, with the candidates chosen from Str\"omgren photometry
(\citealt{Hilker20}). The slit width was fixed to 1$\arcsec$ and the
slit length ranged from 5 to 7$\arcsec$. The seeing was
always better than 1$\arcsec$.  While \cite{Hilker04} and
\cite{Kayser05} examined the main-sequence turn-off and SGB stars,
the present work includes observations of the red giant and upper
main sequence stars in both $\omega$ Cen and M\,55.  Each mask was observed
through a `blue' grism (ESO 600I+25, second order) and a `red' grism
(ESO 1400V+18) with maximum wavelength coverages of 3690 to 4888 \AA\
and 4560 to 5860 \AA . The corresponding dispersions are 0.58 \AA\
pix$^{-1}$ and 0.62 \AA\ pix$^{-1}$, respectively. The exposure times
were chosen in such a way as to ensure that the minimum signal to
noise ratio (S/N) per pixel was $\sim$ 40--50 for the majority of
objects. For giant stars, the average S/N was slightly larger, about
$\sim$ 70--80. In the same way, the observations for M\,55 were carried
out with different masks for main-sequence, subgiant and giant
stars. In order to have reference spectra from stars with known
stellar parameters, long slit spectra with the same grism/filter
combinations were obtained for eleven bright stars with a typical S/N
of about 100--150. These stars are referred to as `standard stars' in
the following. In addition to the science frames, bias, flat field and
wavelength calibration images were obtained during daytime.  The log
of the observations reduced for this work is shown in Table
\ref{tab}. A summary and outline of the reductions and analysis based
on line indices for the main sequence turn-off stars is found in
\cite{Hilker04} and \cite{Kayser05}. It should be noted that the
sample of observed stars is not complete. The focus of these
observations was to get a large sample of main-sequence turn-off and
subgiant stars which are best suited to break the age-metallicity
degeneracy. 
The upper main-sequence and red giant stars primarily serve to provide
additional constraints for isochrone fitting of the different
(age/metallicity) populations in $\omega$ Cen.

\begin{table*}[]
\caption{\label{tab} Table of the observations reduced and analysed
 for this work in addition to those reported by \cite{Kayser05}. The
 different field names in $\omega$ Cen are abbreviated by their
 location in the cluster (e.g om-\textit{e} corresponds to
 \textit{e}ast). The ending `rgb' refers to red giant stars, `to' to
 main-sequence turn-off stars, `ms' correspondingly to upper
 main-sequence stars. The field of view of a MXU mask is 6$\farcm$8
 $\times$ 6$\farcm$8.}  \centering
\begin{tabular}[l]{clcccc}
\hline
 date & mask & $\alpha_{2000}$ & $\delta_{2000}$ & exp.time (red) [sec.] & exp.time (blue) [sec.]\\
 \hline
2002/05/10 & om-n-rgb & 13:26:46.7 & -47:18:43.3 & 1 $\times$ 120,\ 1 $\times$ 240 & 2 $\times$ 300 \\
2002/05/11 & om-e-rgb & 13:27:48.1 & -47:27:27.4 & 2 $\times$ 180 & 2 $\times$ 240 \\
2002/05/11 & om-s1-rgb & 13:26:57.4 & -47:37:43.1 & 2 $\times$ 180 & 2 $\times$ 240 \\
2002/05/11 & om-n-ms & 13:26:46.7 & -47:18:43.2 & 3 $\times$ 1080 & 3 $\times$ 1200 \\
2002/05/12 & om-s2-rgb & 13:26:57.3 & -47:43:42.9 & 2 $\times$ 180 & 2 $\times$ 240 \\
2002/05/12 & om-s2-ms & 13:26:57.3 & -47:43:40.0 & 3 $\times$ 960 & 3 $\times$ 1200 \\
2002/05/12 & om-s1-ms & 13:26:57.8 & -47:37:43.7 & 3 $\times$ 900 & 3 $\times$ 1200 \\
2002/05/12 & om-e-ms & 13:27:48.1 & -47:27:30.8 & 3 $\times$ 900 & 3 $\times$ 1200 \\
 \hline
2002/05/10 & M\,55-to & 19:39:59.1 & -30:53:01.2 & 3 $\times$ 600  & 3 $\times$ 720 \\
2002/05/11 & M\,55-rgb & 19:39:58.9 & -30:52:56.4 & 2 $\times$ 180 & 2 $\times$ 240 \\
2002/05/12 & M\,55-ms & 19:39:58.9 & -30:52:59.0 &   2 $\times$ 900 & 2 $\times$ 900 \\
 \hline
\end{tabular}
\end{table*}

\begin{table*}[]
\caption{\label{stantab} Summary of the stars used for validating the
network model. The \feh , \logg\ and \teff\ values given are the
averages from the works of several authors found in the literature
(see \citealt{Cayrel01}). The uncertainties are either the standard
deviations about the mean values (in the case that multiple values are
listed) or the errors stated by the author (when there is only a
single measurement). The last column shows the number of separate
determinations in each case.}  \centering
\begin{tabular}[l]{lccccccc}
\hline
 & &  & & & &  \\[-3mm]
ID & $\overline{\mbox{\feh
}}$ [dex] & $\sigma$(\feh) [dex] & $\overline{\mbox{\logg
}}$ [dex] & $\sigma$(\logg)[dex] & $\overline{\mbox{\teff
}}$ [K] & $\sigma$(\teff) [K] & num \\ 
\hline
BD-032525$^{*}$ & $-$1.90 & 0.08 & 4.00 & 0.57 & 5739 & 40 & 2\\
BD-052678 & $-$1.89 &  $-$   & 4.43 &  $-$   & 5570 & $-$ & 1\\
BD-084501 &  $-$1.59 & 0.27 & 4.00 &  $-$   & 5750 & 210 & 2\\
BD-133442 &  $-$2.95 & 0.21 & 3.92 & 0.25 & 6240 & 75 & 3\\
HD089499$^{**}$  & $-$2.14 & 0.08 & 2.92 & 0.88 & 4875 & 100 & 3\\
HD097916  & $-$1.05 & 0.19 & 4.09 & 0.10  & 6222 & 140 & 5\\
HD179626  & $-$1.26 & 0.04 & 3.70 & 0.06 & 5625 & 40 & 1\\
HD192718  & $-$0.74 & 0.04 & 4.00 & 0.06 & 5650 & 40 & 1\\
HD193901  & $-$1.08 & 0.12 & 4.38 & 0.32 & 5713 & 70 & 8\\
HD196892  & $-$1.02 & 0.10 & 4.04 & 0.27 & 5900 & 100 & 4\\
HD205650  & $-$1.12 & 0.13 & 3.93 & 0.64 & 5729 & 90 & 3\\ 
\hline
$^{*}$ Spectroscopic binary & & & & & & \\
$^{**}$ Star in binary system & & & & & &  \\
\end{tabular}
\end{table*}

\subsection{Data reduction}
\label{datared}

The data reduction of the observations was done by various tools
within IRAF\footnote{IRAF is distributed by the National Optical
Astronomy Observatories, which are operated by the Association of
Universities for Research in Astronomy, INC., under cooperative
agreement with the National Science Foundation.}. The images of each
mask were bias corrected, (dome) flat fielded and cleaned of cosmic
rays.  
The image was then corrected for pixel-to-pixel variations and in a next step the
individual spectra were extracted, both being performed by the `apall'
task. 
The sky was generally subtracted using adjacent regions in the same
slit, but in a few cases it was necessary to
take the background information from other slit regions in the
mask. In these cases, the sky correction was performed by the task
`skytweak'. Although normally satisfactory, this procedure did result in
erroneous sky subtractions in some cases. We therefore excluded
spectra reduced with `skytweak' from the analysis.  Each
stellar spectrum was wavelength calibrated by 10 to 15 lines of
He, Ne, Hg and Cd and rebinned to a final dispersion of 1 \AA\
pix$^{-1}$, yielding an effective resolution of 2 to 2.5 \AA .

The continuum normalization of the spectra is crucial, since it can
directly influence the parametrization performance.  We tried several
techniques to divide out the continuum. These included different
combinations of line exclusion, median filtering techniques and
continuum fitting functions. It was found that excluding the regions
adjacent to the strongest absorption lines in combination with the
IRAF task `continuum' and careful adjustment of the various parameter
settings therein yielded the best results. But we note
that a good continuum fit is naturally harder to find for metal-rich
stars (\feh\ $\sim$ $-$0.5 dex) due to heavy line blending. This
should be kept in mind when interpreting the parametrization results
for the more metal-rich stars. Furthermore, the S/N in some cases is
not the same for the two grisms for the same object. We did not
explicitly test how this influences the overall parametrization but
the results suggest that this effect is of little importance.

Since the slits cover a range of positions on the mask, the
wavelength coverage is not the same for all stars. In order to produce a
homogeneous grid of all spectra, we restricted the wavelength range
from 3850 to 5700 \AA\ with an overlap of the red and blue grisms'
contributions at 4700 \AA . Moreover, for the analysis only certain
wavelength ranges in the spectra were used, the regions being defined
after extensive simulations on synthetic spectra. These are 3900--4400
\AA , 4820--5000 \AA\ and 5155--5350 \AA . 
As expected, these regions,
found by comparing the parametrization results for networks trained on
different ranges, contain the most prominent metal and hydrogen lines.
The restriction to certain wavelength intervals and the exclusion of
(hopefully) unimportant ranges minimizes the number of inputs to the
network, thus increasing the ratio of training templates to free
parameters in the network (see Sect.\,\ref{networks}).  Examples of
typical spectra are shown in Fig.\,\ref{spectra}.

\section{Grid of observed spectra}
\label{obsspec}

We determined stellar radial velocities by cross-correlating
the spectra with suitably chosen template spectra.  With these data at
hand, we corrected the spectra of each star individually in each grism
(blue and red).  The radial velocities were also used as an extra
indicator for cluster membership (in addition to the position of the
star in the colour-magnitude diagram) by comparing them to the
velocities of the clusters.  These are 175 km s$^{-1}$ and 238 km
s$^{-1}$ for M\,55 and $\omega$ Cen, respectively
(\citealt{Harris96}). In order to account for systematic effects not
corrected in this data reduction (such as small mismatches between
the wavelength calibration and science images), we chose rather large
velocity intervals for the cluster membership decision, namely
[100:200] km s$^{-1}$ for M\,55 and [150:350] km s$^{-1}$ for $\omega$
Cen.  In this way about 6 \% of all stars were excluded as being
nonmembers.

The distributions of the velocity errors from the red and blue grism
are shown in Fig.\,\ref{histerr}. Each error combines the individual
cross-correlating error for a given spectrum (as determined from the
IRAF routine `fxcor') and that of the corresponding template stars.

Some additional spectra were eliminated because of the presence
of light polluting bright stars next to the slit. 
However, two effects coming from bright stars that are close to the
observed field can cause erroneous sky subtraction. First, there may
be reflections that cause gradients in the background, and second,
there may be a contamination in the wings of the PSF. Finally, some
low quality spectra were also eliminated.

We split our observed stars into two luminosity class categories based
on the available Johnson photometry: MSSGB (comprising all
main-sequence and subgiant stars) and RGB (all red giant stars). This
cut is sensible since the stellar parameters \logg\ and \teff\ of
these two groups are significantly different. Moreover, in order to
improve the neural network performance, it is generally recommended to
simplify the problem as much as possible (see
e.g. \citealt{Haykin}). More explicitly, by splitting the
regression problem (the mapping from inputs to outputs) into two parts
(in terms of stellar parameters), we hope that the overall
approximation of the underlying function by the network is improved by
lifting possible ambiguities in the mapping function.  Tests showed
that the results are indeed better when considering two object
categories instead of one. This is certainly also due to the different
S/N of the MSSGB and RGB objects. As described in Sect.\,\ref{dataq},
the giants have generally higher S/N than the subgiants and main
sequence stars. It was therefore necessary to train networks on
different ranges in S/N (see below). For M\,55 there are 130 stars in
the MSSGB and 35 stars in the RGB sample, while for $\omega$ Cen the
numbers are 800 and 66, respectively.

Since we have synthetic training samples only for stars up to
the helium flash, we removed all horizontal branch and suspected
asymptotic giant branch stars in both our M\,55 and $\omega$ Cen
samples. We will consider the parametrization of these in a future
study, which will also include an analysis of $\alpha$ and CN
abundances.

\section{Grid of synthetic spectra}
\label{syndata}

We have calculated synthetic (template) spectra using the model
atmospheres from \cite{Castelli97} in combination with the latest
version of SPECTRUM (\citealt{Gray94}) along with its line list. Since
our study focuses on globular clusters, the metallicity range of the
calculated spectra was limited to $-$ 2.5 dex $\leq$ \feh\ $\leq$ $-$
0.5 dex, while temperature and surface gravity are limited to 3500 K
$\leq$ \teff $\leq$ 8500 K and 0 dex $\leq$ \logg $\leq$ 5.0 dex
respectively. Note that the grid is not complete, i.e. not all
parameter combinations exist. The spectra were calculated with a step
size of 0.02 \AA\ in the wavelength range 3850 to 5700 \AA . The
microturbulence was fixed at 2 km s$^{-1}$. Similarly to the procedure of
\cite{Bailer97}, the spectra were dispersion-corrected and binned
to have the same resolution and wavelength range as the observed data.

As will be described in Sect.\,\ref{networks}, we used spectra with
different S/N to train the network models. To do so, we added the
corresponding noise for a given S/N before normalizing the
spectra. The wavelength of scaling the S/N was chosen at 4500 \AA\
which is in the overlap region of the two grisms.

For the present tests we considered solar scaled $\alpha$ abundances,
something which is empirically found to be only the case for halo
objects with metallicities $> -$1 dex (see
e.g. \citealt{McWilliam97}).  However, this does not pose a problem
since the parametrizer is explicitly trained only on metallicity as
expressed by \feh\ (and not on $\alpha$ abundances). The regression
model (the neural network) is therefore expected to focus on the
changes as caused by different \feh\ values, thus ignoring spectral
dependencies caused by other specific elements.

\section{Artificial Neural Networks}
\label{networks}

Neural networks have proven useful in many scientific disciplines for
analysing data by providing a nonlinear mapping between an input
domain (in this case the spectra) and an output domain (the stellar
parameters). For details on this subject with special emphasis on
stellar parametrization see, for example, \cite{paper1},
\cite{Bailer97}, \cite{synspec}, \cite{Snider2001}, and especially
\cite{Bailer2002} for a general overview. The software used in this
work is that of \cite{soft}.

A neural network can be visualized by several layers, where
each layer is made up of several so-called nodes or neurons. In
general, for the type of network used in this work, one distinguishes
an input-layer, one or two hidden-layers, and an output-layer. The
nodes in each layer are connected to all the nodes in the preceding
and/or following layer via some free parameters (`weights'). While the
nodes in the input layer simply hold the inputs to the network
(i.e. the fluxes in the wavelength bins of the spectrum), each node in
the hidden layer performs a weighted summation of all its inputs. This
sum is then passed through a nonlinear `transfer function' (here the
$tanh$ function) and the result is then passed further to a node in
the next layer. The output nodes do the same as the nodes in the
hidden layer, except that the transfer function is here chosen to be
linear. Before the network can be applied to predict the stellar
parameters for previously unseen (observed) spectra, it must be
trained on an existing set of (here synthetic) spectra with known
stellar parameters. During this training phase, the weights of the
network are iteratively adjusted in order to accurately predict the
outputs (stellar parameters), based on the inputs (spectral
fluxes). This weight adjustment is done according to some measure of
error (`cost function'). Here we use the sum-of-squares error
calculated from the networks' outputs and the corresponding true
stellar parameters for a given synthetic spectrum. Once this error is
sufficiently small, the training is stopped and the weights are
fixed. The network can now be used to predict the stellar parameters
of a star based on the fluxes in a stellar spectrum.

There are several parameters in a neural network which must be
adjusted, often by careful experiments. In the model used here
these are essentially a weight decay term (which inhibits
overfitting by a too complex network), a weighting term for each
stellar parameter in the cost function, the number of hidden layers,
as well as the number of weights in each hidden layer and the number
of training iterations.  After extensive tests we found that a network
with two hidden layers each having 13 neurons provided the best
results. More neurons did not significantly improve the results but
only increased the training time, while a smaller number of
neurons resulted in a poorer model with larger errors.

In principle, one could train three independent network models,
each with a single
output, for each of the three stellar parameters (\teff , \logg\ and
\feh). It is well known, however, that changes in the values of these
stellar parameters may have similar effects on a stellar spectrum
(both lines and continuum). A special example is the helium abundance,
where an increase of this element changes the line shapes in the same
way as does surface gravity (\citealt{Gray92}). To account for such
degeneracies we therefore trained the networks on all three stellar
parameters simultaneously.  
As expected from the above, but contrary to what \cite{Snider2001}
found, a network with multiple outputs performed better than one
trained only on one single parameter. A possible reason for this
might be that we used two hidden layers. It is generally known that a
second hidden layer is important to model the `subtle' information in
the data (here metallicity and surface gravity), see
e.g. \cite{Bishop}. Networks with two hidden layers might therefore be
better suited for this kind of parametrization problem.

\subsection{Network training and regularization}
\label{nettrain}

The observed spectra cover a range of different S/N.  The neural
network model (or the training data) should account for
this. Indeed, we found that networks trained on templates with
high S/N and then validated on spectra with a much lower S/N increased
the bias and (to a smaller degree) also the variance of the
distribution of residuals. This behaviour was also found in tests by
\cite{Snider2001}.  To investigate this further, we trained networks
on different training sets with multiple noise versions of a spectrum
(only one noisy spectrum at each S/N) for a given stellar parameter
combination. In addition, we tried different values for the weight
decay (regularization) parameter $\gamma$. We found that a network
trained only on high S/N templates (S/N =100) can reproduce the
stellar parameters for spectra with different S/N reasonably well, but
only if the weight decay term is chosen to be rather large.  This
shows that this is a regularization problem, i.e. a network trained on
high S/N spectra will overfit the data unless `restrained'. When using
multiple noisy versions of a training template the results improved
slightly, which is understandable since additional noise in the
training data serves as an additional regularization mechanism (noise
being hard to fit). The results of these tests are summarized in
Table\,\ref{nettab}. 

\begin{table}[]
\caption{\label{nettab} Summary of the regularization results
from neural networks with different training sets and weight decay
term $\gamma$. Increasing $\gamma$ results in smaller weights of the
network and thus a more regularized solution. However, too large a value of $\gamma$ will again result in larger deviations, i.e. there is a trade-off in setting this parameter. The metallicity deviations
for the standard stars are given in terms of the median values of the
difference (computed value $-$ literature value). It can be seen that
training on noise-free data and validating on noisy data
systematically underestimated metallicities. The
results demonstrate that noise in the network inputs can help
improve the regularization.}  \centering
\begin{tabular}[l]{ccc}
\hline
 & &    \\[-3mm]
noise in training set & $\gamma$ & \feh\ offset \\
\hline
no & 0.0001 & $-$0.16 \\ 
no & 0.001 & $-$0.11 \\ 
no & 0.01 & $-$0.07 \\ 
yes & 0.0001 & 0.18 \\ 
yes & 0.001 & 0.02 \\ 
yes & 0.01 & 0.05 \\ 
\hline
\end{tabular}
\end{table}

We therefore trained the networks on sets of noisy versions of
the training templates (synthetic spectra). A training sample of size
$N$ is a collection of several stellar parameter combinations
expressed by \textbf{y$_i$} ($i \in [1,N]$) and associated spectral
flux vectors \textbf{x$_i$}, where each of these is represented $P$
times with different S/N ratios, i.e.
\begin{equation}
\left(
\begin{array}{c}
(\boldsymbol{x_{1}};\boldsymbol{y_{1}}) \\
(\boldsymbol{x_{2}};\boldsymbol{y_{2}}) \\
\vdots \\
(\boldsymbol{x_{N}};\boldsymbol{y_{n}}) 
\end{array}
\right)
\mathrm{where}  \,\, \boldsymbol{x_{i}} = \left(
\begin{array}{c} \boldsymbol{x_{i}^{SN_{1}}} \\
\boldsymbol{x_{i}^{SN_{2}}} \\
\vdots \\
\boldsymbol{x_{i}^{SN_{P}}} \\
\end{array}
\right)
\end{equation}
For the MSSGB case (main sequence and subgiants) we used synthetic
spectra with 3.5 dex $\leq$ \logg\ $\leq$ 5.0 dex, 4000 K $\leq$ \teff
$\leq$ 8750 K, $-$2.5 dex $\leq$ \feh $\leq$ $-$0.5 dex and scaled to
S/N ratios between 10 and 100, in steps of 10. For the RGB stars, the
stellar parameters were 0.0 dex $\leq$ \logg\ $\leq$ 3.5 dex, 3500 K
$\leq$ \teff $\leq$ 6000 K and the same metallicity range but with S/N
ratios in the range from 50 to 150, in steps of 10. Note that there is
an overlap in \logg\ and \teff\ for the two samples which is necessary
to properly test those stars with parameters at the edge of the grids.

As mentioned above, the networks can also be tuned by a weighting term
for each stellar parameter in the cost function. We used this term to
set up two network models as follows.

\textit{model1} is a network tuned to yield good metallicity
parametrization results. This network was trained on all three stellar
parameters, i.e. \feh , \logg\ and \teff , but with special weight on
metallicity.

\textit{model2} is a network tuned to yield good \logg\ and
\teff\ performance. As \textit{model1}, this network was trained on
all stellar parameters, but with a higher weight on surface gravity
and temperature.

Note that each model was trained on each of the two training sets
(MSSGB and RGB) separately. Since the minimization process can become
trapped in local minima during learning, we set up a committee of 10
networks (with different initial weight settings) for each of the
models.

As described in Sect.\,\ref{datared}, we used three disjoint
wavelength ranges from the spectrum, yielding a total of 878
wavelength bins (network inputs). Including the `bias' weights in the
input and hidden layers, the total number of weights is therefore
11\,648. This is large and it might be surprising that there are
meaningful results given that there are only 4990 and 4818 training
templates for the MSSGB and RGB cases, respectively. However, the
effective number of free parameters in the network is certainly much
lower, because much of the spectrum contains redundant
information. Furthermore, the smoothness of the input-output function
which is to be approximated by the network will also lower the
effective number of free parameters.

\subsection{Error estimates from Bootstrapping neural networks}

In most applications of neural networks to determine stellar
parameters, uncertainties (error bars) on the estimated parameters are
derived from the distribution of the parameter residuals of some
validation set with known target values. While this method is correct
in order to get an overall estimate of the precision to which
parameters can be determined, it does not yield individual
uncertainties for each determined parameter. This is crucial,
however, since objects that lie close to the boundary of the training
grid tend to have higher uncertainties, given that the performance of
neural networks (or indeed any regression model) is generally weak at
the grid boundaries (see e.g. \citealt{Bishop}).  The concept of using
the bootstrap method for the determination of error bars in the
framework of neural networks is discussed in several articles (e.g.\
\citealt{Heskes97} and \citealt{Dybowski20}). For the special case of
stellar parametrization see \cite{ICAP04}.

Since a neural network is a regression model that maps inputs onto
outputs, we can use the concepts of standard errors and confidence
intervals also for this kind of parametrizer.  There are basically two
sources of uncertainty arising with neural networks. The first stems
from the fact that the training data are noisy and incomplete,
i.e. the construction of a training grid by (randomly) sampling
templates is already prone to sampling variations. The second source
of uncertainty is given by the model limitations which arise from the
optimization failing to find the global optimum.  Another source of
error is an inappropriate (e.g. inflexible) parametrization model
(e.g. linear). Since we assume that our network model approximates the
underlying mapping from the inputs to the outputs reasonably well, the
latter source of uncertainty is not considered further.

The bootstrap was introduced by \cite{Efron79} for estimating various
sample properties such as bias, variance and confidence intervals on
any population parameter estimate.  Given a (random) training sample
\textbf{S} comprising pairs of inputs \textbf{x} and corresponding
outputs $y$, i.e. \textbf{S} =
${(\textbf{x}_1,y_1),(\textbf{x}_2,y_2),...,(\textbf{x}_N,y_N)}$,
drawn from a population $F$, we want to estimate some parameter (e.g.\
the network weights) $\theta = f(F)$. This could be done by
calculating $\hat{\theta} = g(\textbf{S})$ based on \textbf{S}.  The
bootstrap is a data-based method for statistical inference which
allows us to estimate the error in the network outputs from different
values of $\hat{\theta}$.  To determine this bootstrap standard error,
one needs to build bootstrap samples. A bootstrap sample,
\textbf{S}$^{*}$, is a random sample of the same size $N$ as the
original sample which is created by randomly resampling \textbf{S}
with replacement (i.e. \textbf{S}$^{*}$ $\subseteq$ \textbf{S}
$\subset$ $F$). In this way, one obtains $B$ bootstrap samples, the
$b$th bootstrap sample given by
${(\textbf{x}_1^{*b},y_{1}^{*b}),(\textbf{x}_2^{*b},y_2^{*b}),...,(\textbf{x}_N^{*b},y_N^{*b})}$. For
each sample, we minimize $\sum_{n=1}^{N}[y_n^{*b} -
y(\theta;\textbf{x}_n^{*})]^2$ (i.e. we derive $B$ regression
functions by training a network on each of the $B$ bootstrap samples)
to yield $\hat{\theta}^{*b}$ in each case. The (nonparametric)
estimate of the bootstrap standard error, {\sc bse}, is the square
root of the variance of the distribution, which for the $n$th
predicted value $y_{n}$ is

\begin{equation}
{\rm \sc{BSE}} (\textbf{x}) = \sqrt{ \frac{1}{B-1} \sum_{1}^{B}[y(\textbf{x}_n;\hat{\theta}^{*b})-y_{\rm boot}(\textbf{x}_n;\cdot)]^2}
\label{g1}
\end{equation}

\noindent
where $y_{\rm boot}(\textbf{x}_n;\cdot)$ is shorthand for the
bootstrap committee's prediction given by $\frac{1}{B}\sum_{b=1}^{B}
y(\textbf{x}_n;\hat{\theta}^{*b})$.

Note that the random resampling is done over the stellar parameters
and not over the noise versions.

It is generally recommended that the number of bootstrap samples be in
the range from 25 to 200 (\citealt{Efron93}). For this work, we chose
$B$ = 30 to save computational power. Limited tests showed that larger
numbers did not significantly change the results. The random
resampling was done with a uniform random number generator as given in
\cite{NR}.

\begin{table*}
\caption[]{Table of Str\"omgren observations of M\,55. The columns
show the central coordinates of the observed fields and the exposure
times in the different filters. Certain parts of these fields overlap
with those of the FORS2 spectroscopic observations.  }
\label{photab}
\small
\begin{tabular}{ccccccc}
\hline\noalign{\smallskip}
date & field & $\alpha_{2000}$ & $\delta_{2000}$ & exp.time($v$) [sec] & exp.time($b$) [sec] & 
exp.time($y$) [sec] \\
\hline
2001/07/12 & M\,55-1 & 19:40:19.0 & -30:59:30.0 & 2 $\times$ 30, 3 $\times$ 240 & 1 $\times$ 20, 2 $\times$ 90 & 3 $\times$ 10, 3 $\times$ 60 \\
2001/07/13 & M\,55-2 & 19:40:19.0 & -30:53:30.0 & 1 $\times$ 30, 2 $\times$ 240 & 1 $\times$ 20, 2 $\times$ 90 & 1 $\times$ 10, 2 $\times$ 60 \\
2001/07/14 & M\,55-3 & 19:39:47.1 & -30:56:01.1 & 1 $\times$ 30, 2 $\times$ 240 & 1 $\times$ 20, 2 $\times$ 90 & 1 $\times$ 10, 2 $\times$ 60 \\
\hline
\end{tabular}
\end{table*}

\section{Validation of the network model}
\label{modelval}

In the following, the results of the network validation are
presented. As mentioned before, the metallicity results were obtained
from network \textit{model1} which was tuned to yield accurate
results for this parameter. The results for surface gravity \logg\ and
temperature \teff\ were found from network \textit{model2}, i.e. a
network trained on all three stellar parameters as outputs but with a
higher weight on \logg\ and \teff\ (see Sect.\,\ref{nettrain}).

Unless stated otherwise, we summarize the overall parametrization
results in terms of the average absolute deviation (over some set of
spectra), i.e.
\begin{equation} \label{A} A = \frac{1}{N} \cdot \sum_{p=1}^N \left|C(p) -
T(p)\right| \end{equation} 
where $p$ denotes the $p^{\rm th}$ spectrum
and $T$ is the target (or `true') value for this parameter. The
quantity $C(p)$ is the parametrization output averaged over a
committee of 10 networks (see Sect.\,\ref{networks}).

\subsection{Results for M\,55}

In order to compare the networks' outputs of surface gravity and
temperature based on the observed spectra, we have to get an estimate
of the `true' values for these parameters. For this, we did not use
the pre-imaging $BV$ Johnson photometry (because of missing
photometric standard star observations and saturated giant stars), but
rather decided to use more precise measurements of M\,55 obtained in
the Str\"omgren system.  These data were obtained (in addition to
observations in other filters) at the Danish 1.5 m telescope at La
Silla on the nights 12-15 July 2001.  The instrument's field of view
is 6$\farcm$3 $\times$ 6$\farcm$3, the seeing was typically about
1$\farcs$5. A summary of the data reduction is given in the next
section.

\subsubsection{Str\"omgren photometry of M\,55}

Initially the photometric calibration, relating instrumental to
standard magnitudes, was determined. For this we used observations of
12 E region stars (\citealt{Jonch93}) which were reduced in a standard
way (bias subtracted and flat fielded) and the instrumental
magnitudes determined via aperture photometry.  The extracted
magnitudes were then combined in one calibration equation that takes
the airmass of the images into account.  The science images for a
given filter were bias subtracted and flatfielded in a standard
way. The corrected science frames were then averaged to increase the
signal to noise ratio. Instrumental magnitudes were derived via PSF
photometry with the aid of the packages DAOPHOT and ALLSTAR
(\citealt{Stetson87}). A summary of the observations is given in Table
\ref{photab}.  The calibrated data were finally extinction corrected
using the transformation \eby = 0.7 $\cdot$ \ebv\
(\citealt{Crawford70}) and \evy = 1.68 $\cdot$ \eby\ as calculated
from a synthetic extinction curve taken from \cite{Fitzpatrick} in
combination with the Str\"omgren filter transmission curves. Upper
limits for the photometric errors are 0.02 mag for the $y$ and $b$
bands and $0.03$ mag for the $v$ band.  Since the fields of the MXU
and the Danish 1.5m observations only overlapped in certain parts,
photometric information was not available for all objects.

\subsubsection{Surface gravities and temperatures for stars in M\,55} 
\label{secLT55}

To determine \logg\ and \teff\ from the Str\"omgren photometry, we
measured the distances in the three-dimensional $v$,$b$,$y$ space
between the observed magnitudes and the corresponding filter
magnitudes of an appropriate isochrone. For this we used the
isochrone taken from \cite{Bergbusch92} (transformed to Str\"omgren
colours by \citealt{Grebel95}) with an age of 16 Gyr and \feh\ =
$-$1.80 dex with an  assumed cluster distance of 5.3 kpc and a
reddening of \ebv\ = 0.08 mag (\citealt{Harris96}).  We found that
these parameters could best fit the ($b-y$):$y$ and ($v-y$):$y$ colour
magnitude diagrams. The reason why we used Str\"omgren isochrones
which are still on the old age scale is that we had no access to newly
calculated models in this photometric system.

\begin{figure}[t]
\centering
\includegraphics[scale=0.9]{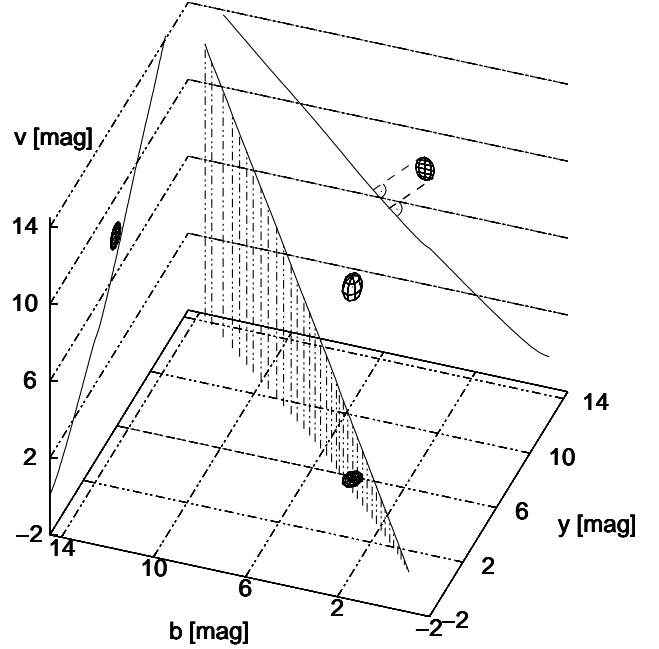}
\caption{A sketch to help visualize the procedure to find
\logg\ and \teff\ values from the Str\"omgren photometry. The solid
line is an isochrone in the three-dimensional magnitude space, while
the ellipsoid is a three-dimensional error box about the measured
point. The size of the error ellipsoid has been magnified by a factor
15 and displaced for clarity.  The surface gravity and temperature for
the measured point is found from the shortest distance between this
point and the isochrone. The errors for these two stellar parameters
are found by projecting the error ellipsoid onto the isochrone and
taking the difference between the extreme values of \teff\ and of
\logg . This is shown in the two-dimensional projection on the $b,v$
plane.  }
\label{d3plot}
\end{figure}

We then `projected' each observed star onto the isochrone by
calculating the distances between the tabulated values of $v$, $b$ and
$y$ of the isochrone and the corresponding observed magnitudes. A star
was then assigned the \logg\ and \teff\ value of the isochrone for
which the calculated distance in this three-dimensional magnitude
space is minimum. Stellar parameter uncertainties $\Delta$(\logg) and
$\Delta$(\teff) are found by defining a (three-dimensional) error
ellipse from the photometric errors $\Delta(v)$, $\Delta(b)$ and
$\Delta(y)$ and by calculating the distance to the isochrone for equally
spaced steps on this ellipsoid.  The errors of \logg\ and \teff\ are
then found from the resulting maximum and minimum values for each
parameter.  This procedure is visualized in Fig.\,\ref{d3plot}.  Note
that because of the slope of the isochrone and the position of a
particular star in the colour magnitude diagram, these errors are not
symmetric.

\begin{figure}[t]
\centering
\includegraphics[scale=0.8]{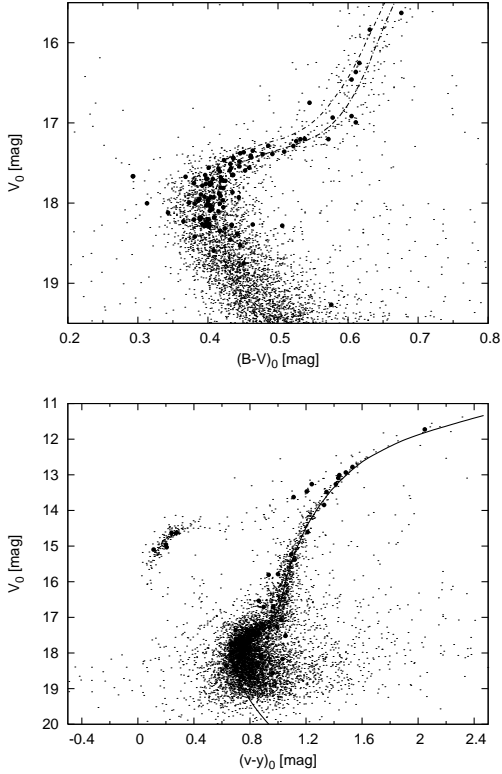}
\caption{Photometry in M\,55. The upper plot is the pre-imaging
Johnson $BV$ photometry for a selected range of luminosities.  Two
isochrones both with 13.5 Gyr and \feh\ = $-$1.8 dex are
overplotted. The left one has solar scaled $\alpha$ abundances, the
right one has [$\alpha$/Fe] = 0.3 dex. The lower plot shows $v-y$
Str\"omgren photometry and an isochrone with \feh\ = $-$1.8 dex and 16
Gyr (the old age scale). The heavy points show the stars for
which spectra have been obtained.
\label{55hrd}}
\end{figure}

Note that the age value of the Str\"omgren isochrone is offset by 2.5
Gyr from the 13.5 Gyr as determined by the best fitting
($\alpha$-enhanced) isochrone for Johnson photometry 
(Fig.\,\ref{55hrd}).  To check for consistency, we did tests with the available
pre-imaging  Johnson photometry 
and the [$\alpha$/Fe] = 0.3 dex Johnson isochrone (taken from
\citealt{Kim02}). We found that there were significant differences (of
about $\sim$ 80 K) in the \teff\ values only for the upper main
sequence and turn-off region stars, while the \logg\ differences were
of the order of 0.1 to 0.2 dex. Since we only want to demonstrate the
principal capability of the networks to derive surface gravity and
temperature estimates, we did not investigate this further.

\begin{figure}[t]
\centering
\includegraphics[scale=0.45]{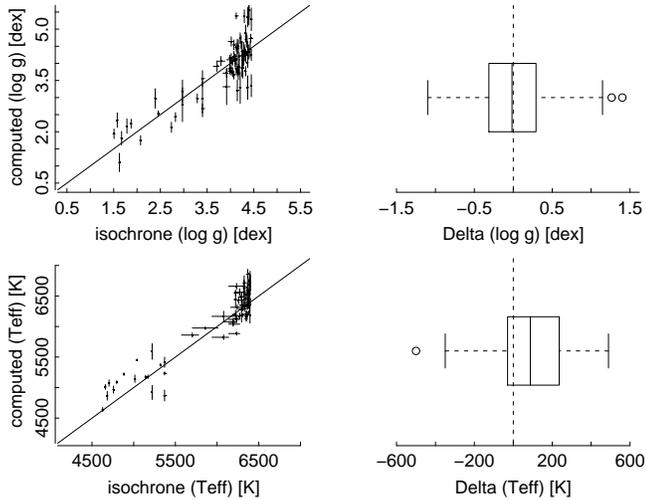}
\caption{Parametrization results for \logg\ (upper line) and \teff\
(lower line) for stars in M\,55. The left panels shows the parameter
values as determined from the isochrone versus the corresponding
computed value from the network for the two parameters. The solid line
is the identity.  The horizontal and vertical error bars for each
point are calculated from the photometric error ellipse (see
Fig.\,\ref{d3plot}) and from 30 bootstrap networks, respectively. The
right panel shows the boxplots of the residuals for each parameter,
$Delta$ = (computed value - isochrone value). The solid line in the
box shows the median value of the residuals while the borders of the
box are the first and third quartiles.
Single points which have a larger distance to the median than 1.5
times the boxlength and are identified as outliers. The overall
parametrization errors as defined in Eq.\,\ref{A} are 0.32 dex and 190
K for the two parameters.
\label{LT55}}
\end{figure}

The parametrization performances for \logg\ and \teff\ for stars in
M\,55 for which Str\"omgren photometry was available are presented in
Fig.\,\ref{LT55}. Note that these results were obtained from network
\textit{model2}, i.e. networks trained with a higher weight on \logg\
and \teff.  In order to appreciate the results one should keep in mind
that the networks were trained on synthetic spectra with a single
microturbulence value of $\xi$ = 2 kms$^{-1}$. However, as
\cite{Gray01} showed, a range of different microturbulences is
important in order to correctly estimate surface gravity values.  It
should also be noted that the \teff\ and \logg\ values as estimated
from the isochrone have uncertainties of their own. This is especially
the case for those stars that lie close to the main-sequence
turn-off, where a perpendicular projection of a star onto the
isochrone (see above) gives rise to larger uncertainties.

From the plots, it can be seen that the parametrization yields
reasonable results. The scatter is somewhat larger for high \logg\ and
\teff\ values, which is most likely due to the fact that the values
determined from the isochrone are uncertain in these regimes. For
\teff\ we observe a small systematic offset of about 80 K.  Given the
intrinsic uncertainties of the target values and the fact that these
are only limited tests (i.e. we did not put too much effort into finding
the optimum network configurations), we conclude that the results are
close to what can be obtained from such data in general
(see Sect.\,\ref{eval}).

\begin{figure}[t]
\centering
\includegraphics[scale=0.3,angle=-90]{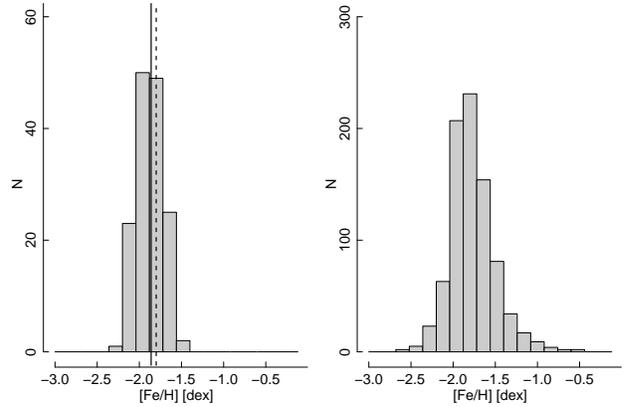}
\caption{Distribution of the metallicities for stars (combined MSSGB
and RGB sample) in M\,55 (left panel) and $\omega$ Cen (right panel)
as determined by our method. The dashed line is centred at $-$1.80
dex (the literature value of \feh\ for M\,55) while the solid line
denotes the mean of the distribution. There is only a small offset of
0.06 dex. The overall error for the determined metallicities in M\,55
as defined in Eq.\,\ref{A} is 0.15 dex.
\label{allF}}
\end{figure}

\subsubsection{Metallicity results for M\,55}

The metallicity values determined from the neural networks for the
combined MSSGB and RGB sample are shown in the left panel in
Fig.\,\ref{allF}. These results were obtained from network
\textit{model1}, i.e. networks trained on all stellar parameters but
with a higher weight on metallicity, see Sect.\,\ref{nettrain}.  The
overall parametrization performance is good.  The mean value of the
metallicity distribution is $-$1.86 dex compared with a literature
value of $-$1.80 dex. The overall error as defined in Eq.\,\ref{A} is
0.15 dex.

\begin{figure}[t]
\centering \includegraphics[scale=0.45]{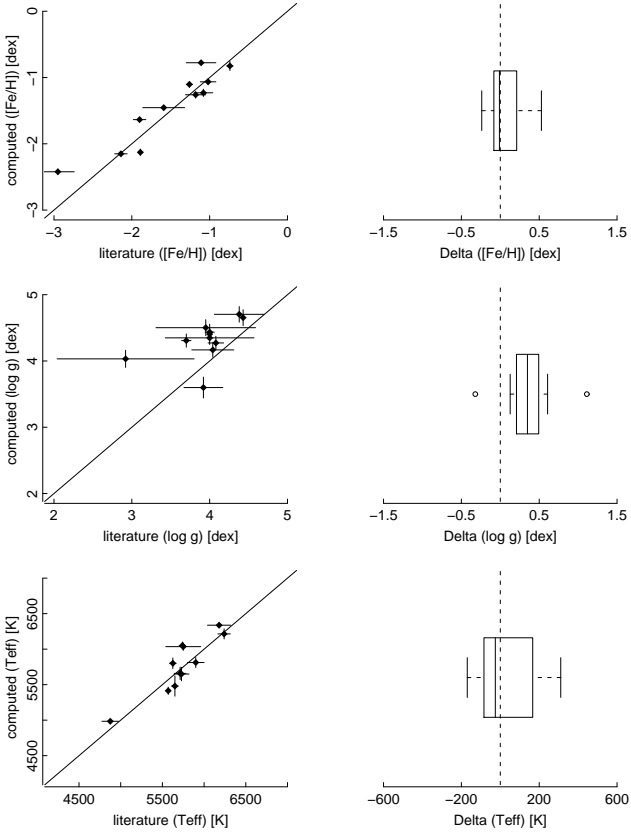}
\caption{As in Fig.\,\ref{LT55} but for the `standard' stars showing the
network outputs compared to the averaged literature values for each of the
parameters \feh , \logg\ and \teff\ (from top to bottom). The errors
for the literature values are found from the scatter about the
unweighted mean value for the different stated values (see
Tab.\,\ref{stantab}), while the computed errors are found from 30
bootstrap networks. Note that the metallicity range on which the
networks were trained was limited to values larger than $-$2.5 dex,
so we cannot expect good parametrization performances for
metallicities below this value. Excluding the single point with the
literature value \feh\ = $-$2.95 dex, we find an overall average error
in \feh\ as defined in Eq.\,\ref{A} of 0.15 dex, while that for \logg\
and \teff\ are 0.42 dex and 145 K respectively.
\label{STANFLT}}
\end{figure}

\subsection{Model validation for standard stars}

We compared the parameter determinations of the neural networks on
standard stars with those found from other methods and listed in the
catalogue of \cite{Cayrel01}. In cases of multiple observations (from
different authors) in this catalogue, we calculated the (unweighted)
mean and the corresponding standard error for those values which were
determined from high resolution spectroscopy. A summary of the values
is given in Table \ref{stantab}. The network models were then
validated on the spectra of these stars (see Sect.\,\ref{dataq}).
We only used the networks trained for the MSSGB sample for this test
(see Sect.\,\ref{obsspec}). The results for \logg\ and \teff\ were
obtained from network \textit{model2} with a higher weight (during
minimization) on surface gravity and temperature, while the results
for \feh\ are from network \textit{model1} with a higher weight on
metallicity, see Sect.\,\ref{nettrain}.  The parametrization results
are summarized in Fig.\,\ref{STANFLT}.

From the plots we see that the metallicity results are quite good,
with an overall error of less than 0.15 dex. Interestingly, the
overall parametrization performance for \logg\ is not as good as for
the stars in M\,55, despite the fact that the S/N of the standard
stars is much higher (but there are only 11 stars so the statistics is
rather poor).  Given that the stellar parameters of the
standard stars are similar to those of the main-sequence stars in
M\,55, and from the fact that the input spectral ranges to
the networks are identical for both types of stars, we conclude that
these larger deviations are, paradoxically, related to the
larger S/N. As outlined above, the S/N of the template spectra with
which the networks were trained influences the degree of
regularization, thus effecting the extent of systematic offsets (see
Sect.\,\ref{networks}). Indeed, the S/N of the template spectra was in
the range from 10 to 100, i.e. lower than that of the standard
stars. To investigate this further, we performed limited tests with networks
trained on spectra with S/N values in the range from 100 to 150 while
keeping the weight decay parameter fixed. We found that the systematic
offset in \logg\ for the standard stars could indeed be decreased in
this way (albeit not completely removed), while the variance remained
almost the same. While these results support the idea of
insufficient regularization, such a strong sensitivity of the network
training to the S/N of the template spectra is surprising and needs to
be investigated in more detail in future studies.  An
explanation of why there are no systematic trends for \feh\ is possibly
given by the fact that the networks for this parameter were highly
tuned, in order to eliminate inaccuracies.  

\begin{figure*}[t]
\centering
\includegraphics[scale=0.9]{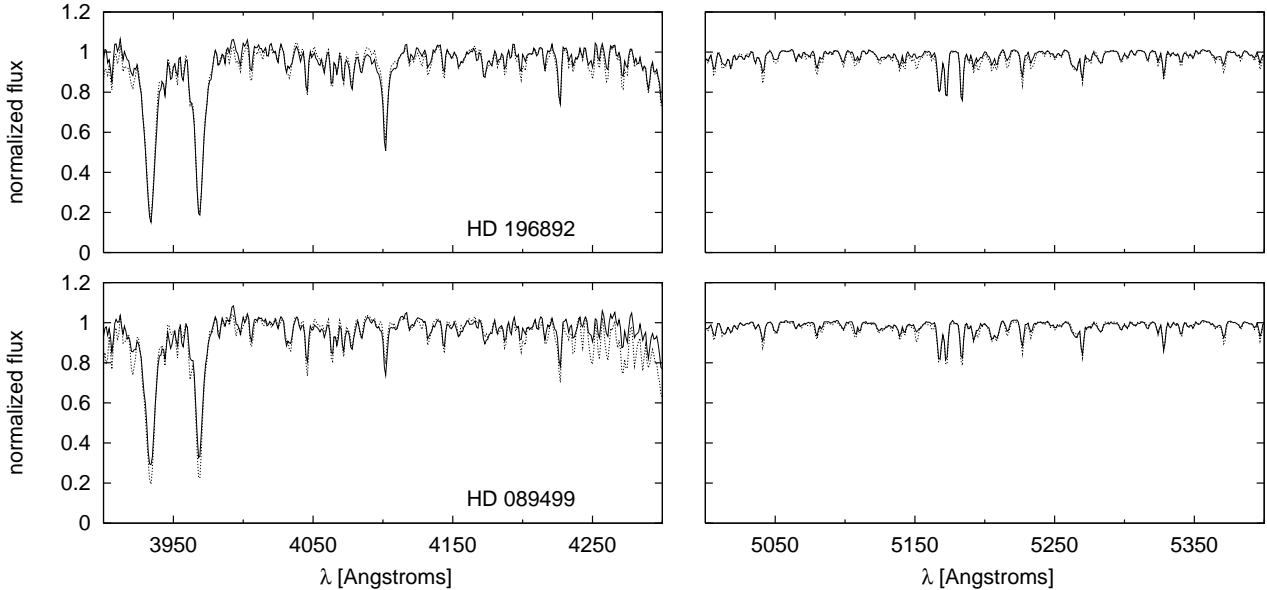}
\caption{A comparison of two observed standard star spectra
(solid lines) with the corresponding synthetic spectra (dotted lines)
as predicted by the neural network. The network predicts only the
stellar parameters, but from these the synthetic spectra were
generated through interpolation of the training template grid of
synthetic spectra. The upper plot is for a star with stellar parameter
residuals of $\Delta$(\feh)= $-$0.04 dex, $\Delta$(\logg)= 0.12 dex and
$\Delta$(\teff)= $-$89 K, where $\Delta$ is the difference (network
output $-$ literature value). The corresponding network outputs are \feh\ = $-$1.06 dex, \logg = 4.16 dex and \teff = 5811 K. The lower plot shows an example were the
difference between the \logg\ prediction of the network and the
literature value is rather large. For this star, we find
$\Delta$(\feh)= $-$0.01 dex, $\Delta$(\logg)=1.11 dex and
$\Delta$(\teff)= 109 K and networks outputs of \feh\ = $-$2.15 dex, \logg = 4.03 dex and \teff = 4984 K. The S/N of the synthetic spectra was decreased
to 150, i.e. similar to that of the observed spectra. 
\label{vergleich}}
\end{figure*}

 Taking the stellar parameters predicted by the network for a
given observed spectrum, we can generate the corresponding synthetic
spectrum. Fig.\,\ref{vergleich} shows a comparison between two
(observed) standard star spectra and the appropriate synthetic
spectrum produced by interpolating the training sample using the
stellar parameters as predicted by the neural networks. It can be seen
that the model spectra represent the observed spectra rather well,
especially in the case that the network's predictions of the stellar
parameters are close to the literature values.  

\begin{figure}[t]
\centering
\includegraphics[scale=0.4]{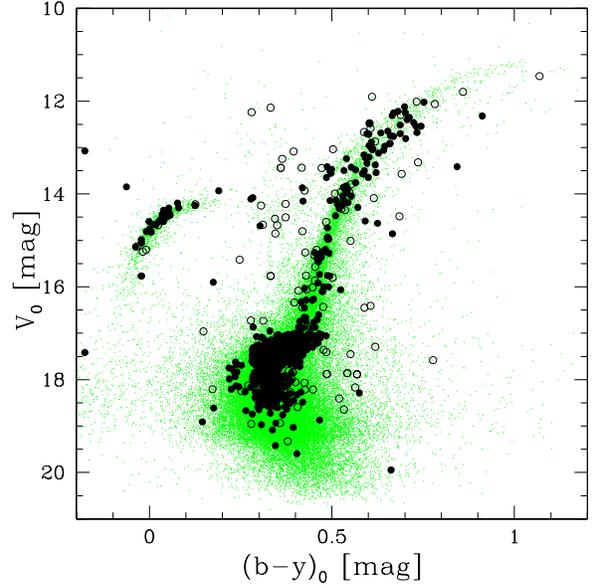}
\caption{The spectroscopically observed stars in $\omega$ Cen are
marked as heavy dots in the Str\"omgren colour magnitude diagram from
\cite{Hilker20}. Based on the radial velocities, filled circles were
identified as members, while open circles denote non-members.
\label{OMhrd1}}
\end{figure}

\subsection{Discussion of the parametrization results}
\label{eval}

The overall parametrization performance can be compared with that of
\cite{Snider2001}. They trained networks on real spectral data with
similar resolutions and for stars with similar parameter ranges as in
this project. They reported precisions of the parameter determinations
for \logg\ of 0.25 to 0.30 dex and 135--150 K for \teff\ while that
for \feh\ was 0.15 to 0.20 dex. Our results are in
absolute agreement with these values, even though our networks were
trained on synthetic data.

\subsection{Metallicity results for $\omega$ Cen}
\label{omapply}

We applied the neural networks to the MSSGB and RGB samples of
$\omega$ Cen to determine metallicities for these stars. We stress,
however, that the metallicities found are somewhat preliminary and we
do not attempt to give a more detailed analysis.  The complex
formation and evolution history of $\omega$ Cen will ultimately
require a much more sophisticated approach in which synthetic spectra
with different helium and/or $\alpha$ element abundances are taken into
account.  The results were obtained from network \textit{model1}
which was tuned to improve metallicity performance.  A plot of all
stars observed is shown in Fig.\,\ref{OMhrd1}. Note that we only
considered the stars up to the top of the giant branch in this study
(see Sect.\,\ref{obsspec}), i.e. there are no HB or AGB stars.  These
will be examined in a future study.

The overall distribution of the metallicities found by our neural
network analysis for $\omega$ Cen is shown in the right panel of
Fig.\,\ref{allF}. When compared to the distribution of metallicities
found for M\,55, we note a significantly larger spread. 
\textit{ From the bootstrap errors for the metallicity values plus the
smaller overall metallicity spread found for the stars in M\,55
(Fig.~\ref{allF}), we must conclude that this large spread does not
indicate an erroneous parametrization but rather a real effect of
the multiple stellar populations in $\omega$ Cen.}

\section{Concluding Remarks}
\label{con}

We have shown that the metallicity information in medium resolution
spectra can be readily assessed by neural networks trained on
synthetic spectra and without additional photometric
information. The neural network approach works in an objective
way and is, ultimately, time efficient, relevant when large sets of
spectra of stars in different objects (clusters) are to be analysed.
In order to obtain some measure of uncertainty on the stellar
parameters determined, we applied the bootstrap method to neural
networks for the first time for this kind of parametrization problem.

The estimated metallicity accuracies are of the order of
$\sim$ 0.2 dex which is close to what one can expect from such data
with any method. We additionally performed limited tests on the retrieval
of \logg\ and \teff\ and found that the overall accuracies for these
parameters are $\sim$ 0.3 to 0.4 dex and $\sim$ 150 to 190 K. These
results could probably be further improved once a suitable grid of
synthetic spectra with a range of different microturbulent velocities
is available. In addition, the effects of $\alpha$ abundances should
be taken into account to test how well this additional parameter can
be determined and whether neural networks are able to discriminate between the
effects of \feh\ and [$\alpha$/Fe] from normalized spectra alone.
Moreover, and especially with regard to $\omega$ Cen, the recent
suggestions of enhanced helium abundances for the intermediate
metallicity population (\citealt{Norris04}) should be taken into
account in future simulations of the synthetic templates. As
mentioned above, an increased helium abundance results in an overall
increased molecular weight which affects the absorption lines in
similar ways to \logg .
 
Future tests could also include photometric data (e.g. from the
pre-imaging observations). Since temperature strongly affects the
continuum of a stellar energy distribution, any colour information
should improve the performance for this parameter.  Once a larger set
of observed spectra for stars in normal globular clusters (i.e. with a
well defined metallicity) is available, we will train
networks on these data. Although synthetic spectra can reproduce
observed stellar energy distributions rather well, we believe that a
network trained on real data with similar noise and detector
characteristics will result in better metallicity estimates for
previously unseen data.

\begin{acknowledgements}
PGW is grateful to the DLR for support under the grant 50QD0103.  We
thank R.O. Gray for distributing the SPECTRUM code. We also thank
K.S. de Boer and T.A. Kaempf for helpful discussions, O. Cordes and
O. Marggraf for assistance with the computer system level support and
the referee for helpful comments.
\end{acknowledgements}

\bibliographystyle{apj}
\bibliography{1974.bib}

\begin{thebibliography}{43}
\expandafter\ifx\csname natexlab\endcsname\relax\def\natexlab#1{#1}\fi

\bibitem[{{Bailer-Jones}(1998)}]{soft}
{Bailer-Jones}, C. A.~L. 1998, Statnet - a feedforward interpolation neural
  network, Tech. rep., http://www.mpia-hd.mpg.de/homes/calj/statnet.html

\bibitem[{{Bailer-Jones}(2000)}]{synspec}
{Bailer-Jones}, C. A.~L. 2000, A\&A, 357, 197

\bibitem[{{Bailer-Jones}(2002)}]{Bailer2002}
{Bailer-Jones}, C. A.~L. 2002, in Automated {D}ata {A}nalysis in {A}stronomy,
  eds. R.~Gupta, H.~Singh, \& C.~A.~L. Bailer-Jones (Narosa {Publishing}
  {House}, {New} {Delhi}, {India}), 51

\bibitem[{{Bailer-Jones} {et~al.}(1997){Bailer-Jones}, {Irwin}, {Gilmore}, \&
  {von Hippel}}]{Bailer97}
{Bailer-Jones}, C.~A.~L., {Irwin}, M., {Gilmore}, G., \& {von Hippel}, T. 1997,
  MNRAS, 292, 157

\bibitem[{{Bergbusch} \& {VandenBerg}(1992)}]{Bergbusch92}
{Bergbusch}, P.~A. \& {VandenBerg}, D.~A. 1992, \apjs, 81, 163

\bibitem[{Bishop(1995)}]{Bishop}
Bishop, C. 1995, Neural Networks for Pattern Recognition (Oxford University
  Press)

\bibitem[{{Bonifacio} \& {Caffau}(2003)}]{Bonifacio03}
{Bonifacio}, P. \& {Caffau}, E. 2003, \aap, 399, 1183

\bibitem[{{Castelli} {et~al.}(1997){Castelli}, {Gratton}, \&
  {Kurucz}}]{Castelli97}
{Castelli}, F., {Gratton}, R.~G., \& {Kurucz}, R.~L. 1997, \aap, 318, 841

\bibitem[{{Cayrel de Strobel} {et~al.}(2001){Cayrel de Strobel}, {Soubiran}, \&
  {Ralite}}]{Cayrel01}
{Cayrel de Strobel}, G., {Soubiran}, C., \& {Ralite}, N. 2001, \aap, 373, 159

\bibitem[{{Crawford} \& {Barnes}(1970)}]{Crawford70}
{Crawford}, D.~L. \& {Barnes}, J.~V. 1970, \aj, 75, 978

\bibitem[{{Dybowski} \& Roberts(2000)}]{Dybowski20}
{Dybowski}, R. \& Roberts, S.~J. 2000, in Clinical Applications of Artificial
  Neural Networks, eds. R.~Dybowski \& V.~Gant (Cambridge University Press)

\bibitem[{{Efron}(1979)}]{Efron79}
{Efron}, B. 1979, Ann. Statist., 7, 1

\bibitem[{Efron \& Tibshirani(1993)}]{Efron93}
Efron, B. \& Tibshirani, R. 1993, An Introduction to the Bootstrap (Chapman and
  Hall, New York)

\bibitem[{{Fitzpatrick}(1999)}]{Fitzpatrick}
{Fitzpatrick}, E.~L. 1999, PASP, 111, 63

\bibitem[{Gray(1992)}]{Gray92}
Gray, D.~F. 1992, The observation and analysis of stellar photospheres
  (Cambridge Astrophysics Series)

\bibitem[{{Gray} \& {Corbally}(1994)}]{Gray94}
{Gray}, R.~O. \& {Corbally}, C.~J. 1994, \aj, 107, 742

\bibitem[{{Gray} {et~al.}(2001){Gray}, {Napier}, \& {Winkler}}]{Gray01}
{Gray}, R.~O., {Napier}, M.~G., \& {Winkler}, L.~I. 2001, ApJ, 121, 2148

\bibitem[{{Harris}(1996)}]{Harris96}
{Harris}, W.~E. 1996, \aj, 112, 1487

\bibitem[{Haykin(1999)}]{Haykin}
Haykin, S. 1999, Neural {N}etworks - {A} {C}omprehensive {F}oundation (Prentice
  {Hall}, {Upper} {Saddle} {River}, {New} {Jersey})




\bibitem[{{Heskes}(1997)}]{Heskes97}
{Heskes}, T. 1997, in {A}dvances in {N}eural {I}nformation {P}rocessing
  {S}ystems, eds. M.~C.Mozer,  M.~I. Jordan, \& T. Petsche (The MIT Press) 
9, 176





\bibitem[{{Hilker} {et~al.}(2004){Hilker}, {Kayser}, {Richtler}, \&
  {Willemsen}}]{Hilker04}
{Hilker}, M., {Kayser}, A., {Richtler}, T., \& {Willemsen}, P. 2004, \aap, 422,
  L9

\bibitem[{{Hilker} \& {Richtler}(2000)}]{Hilker20}
{Hilker}, M. \& {Richtler}, T. 2000, \aap, 362, 895

\bibitem[{{Hughes} {et~al.}(2004){Hughes}, {Wallerstein}, {van Leeuwen}, \&
  {Hilker}}]{Hughes04}
{Hughes}, J., {Wallerstein}, G., {van Leeuwen}, F., \& {Hilker}, M. 2004, \aj,
  127, 980

\bibitem[{{J{\o}nch-S{\o}rensen}(1993)}]{Jonch93}
{J{\o}nch-S{\o}rensen}, H. 1993, \aaps, 102, 637

\bibitem[{{Kayser} {et~al.}(2005){Kayser}, {Hilker}, {Richtler}, \&
  {Willemsen}}]{Kayser05}
{Kayser}, A., {Hilker}, M., {Richtler}, T., \& {Willemsen}, P.~G. 2005, {A}\&A,
  in prep.

\bibitem[{{Kim} {et~al.}(2002){Kim}, {Demarque}, {Yi}, \& {Alexander}}]{Kim02}
{Kim}, Y., {Demarque}, P., {Yi}, S.~K., \& {Alexander}, D.~R. 2002, \apjs, 143,
  499

\bibitem[{{Lee} {et~al.}(1999){Lee}, {Joo}, {Sohn}, {Rey}, {Lee}, \&
  {Walker}}]{Lee99}
{Lee}, Y.-W., {Joo}, J.-M., {Sohn}, Y.-J., {et~al.} 1999, \nat, 402, 55

\bibitem[{{McWilliam}(1997)}]{McWilliam97}
{McWilliam}, A. 1997, ARA\&A, 35, 503

\bibitem[{{Norris}(2004)}]{Norris04}
{Norris}, J.~E. 2004, \apjl, 612, L25

\bibitem[{{Norris} \& {Da Costa}(1995)}]{Norris95}
{Norris}, J.~E. \& {Da Costa}, G.~S. 1995, \apj, 447, 680

\bibitem[{{Origlia} {et~al.}(2003){Origlia}, {Ferraro}, {Bellazzini}, \&
  {Pancino}}]{Origlia03}
{Origlia}, L., {Ferraro}, F.~R., {Bellazzini}, M., \& {Pancino}, E. 2003, \apj,
  591, 916

\bibitem[{{Pancino} {et~al.}(2002){Pancino}, {Pasquini}, {Hill}, {Ferraro}, \&
  {Bellazzini}}]{Pancino02}
{Pancino}, E., {Pasquini}, L., {Hill}, V., {Ferraro}, F.~R., \& {Bellazzini},
  M. 2002, \apjl, 568, L101

\bibitem[{{Press} {et~al.}(1992){Press}, {Vetterling}, \& {Flannery}}]{NR}
{Press}, W. H., {Teukolsky}, S.~A., {Vetterling}, W.~T., \& {Flannery}, B.~P.
  1992, Numerical Recipes in C (Cambridge University Press)

\bibitem[{{Richter} {et~al.}(1999){Richter}, {Hilker}, \&
  {Richtler}}]{Richter99}
{Richter}, P., {Hilker}, M., \& {Richtler}, T. 1999, \aap, 350, 476

\bibitem[{{Roberts} \& {Grebel}(1995)}]{Grebel95}
{Roberts}, W.~J. \& {Grebel}, E.~K. 1995, Bulletin of the American Astronomical
  Society, 27, 816



\bibitem[{{Smith}(2004)}]{Smith04}
{Smith}, V.~V. 2004, in Origin and Evolution of the Elements, eds. A.~McWilliam, M.~Rauch (Carnegie Observatories Astrophysics Series), 188


\bibitem[{{Snider} {et~al.}(2001){Snider}, {Allende Prieto}, {von Hippel},
  {Beers}, {Sneden}, {Qu}, \& {Rossi}}]{Snider2001}
{Snider}, S., {Allende Prieto}, C., {von Hippel}, T., {et~al.} 2001, ApJ, 562,
  528

\bibitem[{{Soubiran} {et~al.}(1998){Soubiran}, {Katz}, \&
  {Cayrel}}]{Soubiron98}
{Soubiran}, C., {Katz}, D., \& {Cayrel}, R. 1998, A\&AS, 133, 221

\bibitem[{{Stetson}(1987)}]{Stetson87}
{Stetson}, P.~B. 1987, \pasp, 99, 191

\bibitem[{{Suntzeff} \& {Kraft}(1996)}]{Suntzeff96}
{Suntzeff}, N.~B. \& {Kraft}, R.~P. 1996, \aj, 111, 1913

\bibitem[{Willemsen {et~al.}(2004)Willemsen, Bailer-Jones, \& Kaempf}]{ICAP04}
Willemsen, P.~G., Bailer-Jones, C.~A.~L., \& Kaempf, T.~A. 2004, Analysis of Stellar
  Parameter Uncertainty Estimates from Bootstrapping Neural Networks, Tech.
  rep., Gaia-ICAP, ICAP-PW-004

\bibitem[{{Willemsen} {et~al.}(2003){Willemsen}, {Bailer-Jones}, {Kaempf}, \&
  {de Boer}}]{paper1}
{Willemsen}, P.~G., {Bailer-Jones}, C.~A.~L., {Kaempf}, T.~A., \& {de Boer},
  K.~S. 2003, \aap, 401, 1203

\bibitem[{{Zinn} \& {West}(1984)}]{Zinn84}
{Zinn}, R. \& {West}, M.~J. 1984, \apjs, 55, 45

\end{thebibliography}

\end{document}